\documentclass[aps,final,notitlepage,oneside,twocolumn,nobibnotes,nofootinbib,
superscriptaddress,noshowpacs,centertags]{revtex4-1}

\usepackage[utf8]{inputenc}
\usepackage[english]{babel}
\usepackage{graphicx}
\usepackage{latexsym}
\usepackage{amssymb}
\usepackage{amsmath}
\usepackage{float}

\begin{document}

\title{Evolution of the DM distribution function in the density spikes around PBHs}
\author{Yu.N. Eroshenko}\thanks{e-mail: eroshenko@inr.ac.ru}
\affiliation{Institute for Nuclear Research of the Russian Academy of Sciences, Moscow 117312, Russia}

\date{\today}

\begin{abstract}
At the cosmological stage of radiation dominance, dark matter density spikes should form around primordial black holes. In the case when dark matter particles are able to annihilate, the density in the central regions of the spikes decreases due to the elimination of particles, which gives an upper bound on the central density. In this paper, the modification of the central density profile is investigated, taking into account the distribution of the particle orbits. The orbits in spike around a primordial black hole are very elongated, almost radial, and the angular momentum distribution has an exponential form. For such an initial distribution function, it is obtained that a cusp with the exponent $\approx-0.7$ is formed in the central region, instead of an annihilation plateau. The presence of the cusp provides some correction to the rate of dark matter annihilation around primordial black holes.
\end{abstract}

\maketitle

\section{Introduction}
 
The formation of primordial black holes (PBHs) in the early universe was proposed in \cite{ZelNov67} and later a similar model was considered in \cite{Haw71}. In the work \cite{CarHaw74}, for the first time, the absence of a catastrophic PBH mass growth due to the accretion of surrounding matter was shown, which gave the PBH hypothesis a greater ground. Gravitational collapses of curvature perturbations were considered in early models, but later several alternative models of PBH formation were proposed \cite{KhlPol0,BerKuzTka83,Jed97,DolSil93,RubKhlSak00,RubSakKhl01}. Although the existence of PBHs has not yet been proven, they can explain a number of phenomena in astrophysics and cosmology, including part of the LIGO/Virgo/KAGRA gravitational wave events \cite{Naketal97,Ioketal98,Sasetal16,Doletal20}. It is possible that PBHs make up some part of dark matter (DM), however, there are only small mass intervals in which PBHs can provide a significant part of DM. For other masses, their fraction in the composition of DM is very small, $f\ll 1$, see the review \cite{CarKuh20}. In this case, the remaining part of DM may be in the form of some elementary particles such as WIMPs. In this paper we assume that the predominant part of DM consists of WIMPs, for example, neutralinos, which are capable of annihilation, and the annihilation cross section corresponds to the thermal mechanism of their birth. 

In a number of works, the process of DM accumulation on orbits around a PBH at the radiation-dominated stage of the universe evolution has been studied. The physics of the DM density spike formation around a PBH is described in detail in \cite{Ero16}. Also in the works \cite{Bouetal18,Adaetal19,CarKuhVis21,Ire24} various approaches to the calculation of the density spike have been developed. Here we will only briefly describe some main points. Until the moment of kinetic decoupling of DM particles from radiation, they were bound to radiation and could not move freely, but followed the general hydrodynamic flows. In particular, they could not accumulate significantly around a PBH, because this was prevented by the pressure of radiation opposing the gravitational force of the PBH. After the moment of kinetic decoupling, the DM particles began to move freely and some of them could remain in finite orbits around the PBH. These particles formed dense spikes around the PBH at the stage of radiation dominance. At low PBH masses, the residual thermal velocities of DM particles remaining after kinetic decoupling are important, but at large PBH masses, the role of residual velocities decreases. After the transition of the universe to the dust-like stage, the formation of the DM halos around PBHs continued on a large scales, and at this epoch the mass of the halo already exceeded the mass of the seed PBH, and the formation of the halo occurred in the regime of secondary accretion \cite{Ber85}. We will consider only the densest inner part of the DM condensation around a PBH, which was formed before the transition to the dust-like stage.

For the first time, the DM density spikes around the PBHs formed at the stage of radiation dominance were considered in the works \cite{LacBea10,SaiShi11,SaiShi11-2,Don11}. It is interesting to note that the basic equations describing the density profile of the density spikes were already written out in the remarkable work \cite{SaiShi11-2} and then independently obtained in subsequent works. In \cite{LacBea10}, the density profile $\rho\propto r^{-3/2}$ arising from the adiabatic growth of a black hole was assumed as the main option, but also the possibility of $\rho\propto r^{-9/4}$ profile was mentioned. Further research has shown that the formation of the $\rho\propto r^{-9/4}$ profile is more likely. In \cite{LacBea10}, the incompatibility of the model with a large number of PBHs with sufficiently large masses and annihilating DM in the form of WIMPs was pointed out for the first time. This is due to an unacceptably large signal from the annihilation.  In \cite{Ero16}, the structure of the density spike was calculated by a more general method, taking into account the residual thermal velocities of DM particles remaining after kinetic decoupling. It was shown that the density spike probably has a more complex shape, rather than a simple power-law one, associated with different modes of its formation at different distances from a PBH. In \cite{Bouetal18}, the calculation method has been significantly improved. In \cite{Adaetal19}, a profile close to $\rho\propto r^{-9/4}$ was obtained by numerical modelling and analytical calculations. In \cite{Adaetal19} it was noted that for PBH masses $M>10^{-6}M_\odot$, the thermal velocities of DM particles can be neglected. In the \cite{Adaetal19} approach, the equation of the radial motion of particles in the PBH gravitational field at the cosmological background played a great role, and in \cite{Ero19} this equation was derived in a self-consistent manner within the general relativity using the velocity field obtained in \cite{BabDokEro18}. Then, in the work \cite{CarKuhVis21}, the issue of compatibility of PBHs and DM in the form of WIMPs was considered once again with a wide range of possible parameters, and new more complete constraints were obtained. Finally, the density profile was also calculated in \cite{Ire24}, where an approximate analytical expression $\rho\propto r^{-9/4}$ was obtained in the limit of zero thermal velocities of DM particles.

To calculate the gamma radiation flux from the DM annihilation, it is important to know the density profile of the spike and the maximum possible density in its central region. The maximum density of DM can be estimated from the simple condition $\rho\langle\sigma_{\rm ann} v\rangle t_0/m\sim1$       
 \cite{BerGurZyb92,SilSte93}, where $\langle\sigma_{\rm ann} v\rangle$ is the annihilation cross section averaged together with velocity, $m$ is the mass of DM particles, $t_0$ is the age of the universe. Such an calculation gives a central plateau with the constant DM density  
\begin{eqnarray}
&& \rho_{\rm max}\simeq \frac{m}{\langle\sigma_{\rm ann} v\rangle t_0} \sim10^{-14}\left(\frac{m}{70\mbox{~GeV}}\right)\times
\label{annwr}
\\
&\times &\left(\frac{\langle\sigma_{\rm ann} v\rangle}{3\times10^{-26}\mbox{~cm$^3$s$^{-1}$}}\right)^{-1}
\!\!\!\left(\frac{t_0}{1.36\times10^{10}\mbox{~yrs}}\right)^{-1}\mbox{~g~cm$^{-3}$}.
\nonumber
\end{eqnarray}
In the plateau region, the DM density should have fallen to the value of $\rho_{\rm max}$ by now. In this approach, it is assumed that density change is a local process occurring independently at each radius. Meanwhile, DM particles can generally move along elongated orbits, and when particles are annihilated in one place, their number changes in other places, at other distances from the center of the halo. In the work \cite{LacBea10}, the lifetime of the DM particle was estimated, taking into account the fact that during its orbital motion the particle passes through regions with different densities. In \cite{LacBea10}, an orbit with only one characteristic value of the orbital parameters was considered. To slightly refine the estimate (\ref{annwr}), the  equation $\partial\rho/\partial t=-\langle\sigma_{\rm ann} v\rangle\rho^2/m$ can be considered, which gives a smooth transition to the plateau region \cite{GonSil99} and is accurate only for the case of circular orbits \cite{Vas07}. However, this approach is not sufficient to calculate the structure of the transition region for elongated orbits. The next improvement, which is the purpose of this work, is to take into account the distribution of DM particles by orbital parameters. It can be expected that the presence of elongated orbits entangling with each other will lead, during particle annihilation, to the formation not of the plateau, but some cusp in the central region. In the work \cite{Vas07}, this was shown for the case of astrophysical black holes with a power-low distribution of angular momentum. In this paper, we investigate this issue for PBHs, for which, as will be shown, the distribution of DM particles over angular moments is exponential. Taking into account the annihilation of DM particles means that the distribution function also depends on time, $f=f(t,E,L^2)$. In some regions of $E$ and $L$ variation, the number of particles decreases faster than in others, resulting in the evolution of the distribution function.  

The solution of the dynamic problem, when annihilation is considered simultaneously with the process of the density profile formation is difficult. But such a detailed consideration is not required in many cases, because the formation of the central DM spike should occur fairly quickly. For PBH, it is still at the stage of radiation dominance, and for astrophysical black holes at the stage of adiabatic black holes growth. At the same time, annihilation is, as a rule, a slower process, and the final annihilation ``plateau'' in the density profile is formed only by now, i.e. at a time interval that may exceed the time of formation of the central peak by several orders of magnitude. In this regard, it is sufficient to consider a simpler problem with the annihilation of DM particles in an already formed spike. The dynamics of the formation of the density spike affects annihilation only in the region close to the black hole, where the particles have already been completely annihilated (for the annihilation cross-section of WIMPs), therefore this central region can be excluded from the consideration. 

The result of our calculation is the shape of the spike in its central part, which is affected by the DM annihilation. We show that the shape differs from a flat plateau and is close to a power-law cusp with the exponent $\approx-0.7$. This difference slightly changes the rate of annihilation in the center of the halo and, accordingly, the predicted flux of annihilation radiation. 

In the Section~\ref{intdifsec} of this article, we present a general method for describing the transformation of the DM particle distribution function due to annihilation. In the Section~\ref{fpbh}, we present a new method for calculating the density spike profile around a PBH using a distribution function. In the Section~\ref{itogsec}, we numerically solve the integro-differential equations from the Section~\ref{intdifsec} for the case of the initial distribution function around the PBH obtained in the Section~\ref{fpbh} and find the final density profile. All numerical values are given for the example when DM particles have masses $m=70$~GeV and annihilation cross section $\langle\sigma_{\rm ann} v\rangle=3\times10^{-26}$~cm$^3$s$^{-1}$. The kinetic decoupling temperature is $T_d=27$~MeV, which corresponds to the moment of cosmological time $t_d=0.001$~s, see \cite{Ero16}. We consider only dark matter with negligibly weak self-interaction or without interaction (except for annihilation). For example, we assume the PBH mass $M=30M_\odot$. The cosmological parameters are taken from the data of the Planck satellite.


\section{Transformation of the distribution function during annihilation}
\label{intdifsec}

Let's write down the kinetic Boltzmann equation for a nonstationary spherically symmetric distribution of DM particles around a PBH, taking into account their annihilation
\begin{equation}
\frac{\partial f}{\partial t}+\frac{\partial f}{\partial \vec r}\dot{\vec r}+\frac{\partial f}{\partial \vec v}\dot{\vec v}=-f\langle\sigma_{\rm ann} v\rangle n(t,r),
\label{bolts}
\end{equation}
where the distribution function $f(t,\vec r,\vec v)$ is normalized so that the number density of particles is given by 
\begin{equation}
n(t,r)=\int fd^3v.
\end{equation}

As mentioned earlier, annihilation is generally a much slower process compared to the orbital period of the particles. Note that when annihilation is ``turned off'', the distribution function ceases to evolve if insignificant transients on the several orbit time scale are neglected.  Therefore, the following two equations must be fulfilled separately 
\begin{equation}
\frac{\partial f}{\partial \vec r}\dot{\vec r}+\frac{\partial f}{\partial \vec v}\dot{\vec v}=0
\end{equation}
and
\begin{equation}
\frac{\partial f}{\partial t}=-f\langle\sigma_{\rm ann} v\rangle n(t,r).
\label{dfdtann1}
\end{equation}
The first of them, as is known, leads to the conclusion that $f(t,\vec r,\vec v)$ depends only on the integrals of motion $E$ (energy per unit mass) and $L^2$ (square of angular momentum calculated per unit mass), if we neglect the slow change over time associated with annihilation. 

Following the standard procedure, using the Jacobian of the transition from variables ($r$, $\vec v$) to ($E$, $L$), we write 
\begin{equation}
n(t,r)=\int\limits_{E_{\rm min}} dE'\int\limits_{L_g} dL'\frac{4\pi L'}{r^2v_r}f(t,E',L'^2),
\label{neq}
\end{equation}
where the minimum angular momentum is determined by the last circular orbit around a Schwarzschild black hole $L_g=2cr_g$ \cite{GonSil99}, $r_g=2GM/c^2$, $E_{\rm min}\approx 0$, and the radial velocity
\begin{equation}
v_r(E',L')=2^{1/2}\left(E'+\frac{GM}{r}-\frac{L'^2}{2r^2}\right)^{1/2}.
\label{vreq}
\end{equation}
The exact value of $L_g$ does not matter, because the particles in the most central region of the density spike are obviously completely annihilated.

\begin{figure}[tbp]
\centering
\includegraphics[angle=0,width=0.49\textwidth]{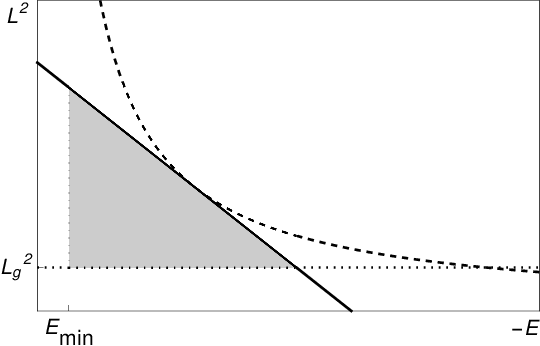}
\caption{Lindblad diagram showing all possible parameters of particle orbits. The integration area in (\ref{neq}) is shaded. The curves are described in the text.}
\label{gr1}
\end{figure}

In the Lindblad diagram, the integration region in (\ref{neq}) is shown at Fig.~\ref{gr1} with a shaded area. Each point of the diagram corresponds to a certain orbit of DM particles. The upper dashed line contains all circular orbits with an eccentricity $e=0$, i.e.
\begin{equation}
L^2=-\frac{G^2M^2}{2E}.
\end{equation}
The solid curve is a characteristic curve for the radius $r$, i.e. orbits where $v_r(r)=0$. The equation of this curve is
\begin{equation}
L^2=2r^2\left(E+\frac{GM}{r}\right).
\end{equation}
The lower horizontal curve shows the minimum square of angular momentum $L^2=L_g^2$.

At any moment, a particle is located at some specific point in its orbit at radius $r$, where, when it encounters particles located there, and it can annihilate. The equation (\ref{dfdtann1}) still has a local form associated with a specific point at the orbit. But, as it was noted in \cite{Vas07,LacBea10}, the particle in its orbital motion passes through regions with different densities, therefore integration along the orbit is necessary. We integrate the equation (\ref{dfdtann1}) along the orbit and obtain the equation of evolution over long times, much large then orbital periods $T_{\rm orb}$. To do this, note that in the range of radii $dr$, the particle spends $2dt/T_{\rm orb}$ part of its time, where $(dt/dr)dr$ can be found from the equation of motion as $dr/v_r$. Therefore, we get
\begin{equation}
\frac{\partial f(t,E,L)}{\partial t}=-\frac{2f\langle\sigma_{\rm ann} v\rangle}{T_{\rm orb}} \int\limits_{r_{\rm min}}^{r_{\rm max}}n(t,r)\frac{dr}{v_r},
\label{dfdtann2}
\end{equation}
where $r_{\rm min}$ and $r_{\rm max}$ are the roots of the equation $v_r=0$, in which $v_r(E,L)$ is expressed in terms of $E$ and $L$ according to the Eq.~(\ref{vreq}), and the orbital period is
\begin{equation}
T_{\rm orb}=\frac{\pi G M}{2^{1/2}|E|^{3/2}}.
\end{equation}

Thus, to study the evolution of the function $f(t,E,L^2)$ the solution of the integro-differential equations (\ref{neq}) and (\ref{dfdtann2}) is required, which in general can only be performed numerically. In \cite{Vas07}, similar equations were obtained and some of their general properties were investigated for the case of power-law dependence $f(t,E,L^2)$ on $E^2$. In this paper, we investigate density spikes around a PBH, where the physical conditions are very different and the dependence is exponential. To solve it, one also needs to set the initial $f(t_i,E,L^2)$ at some initial time $t_i$. As noted earlier, the duration of the density spike formation process around a PBH is much shorter than the duration of the subsequent transformation of the density profile in its central region due to DM particle annihilation, therefore, the exact choice of the moment $t_i$ at $t_i\ll t_0$ is not important.


\section{DM particle distribution function around a PBH}
\label{fpbh}

Now let's calculate the density profile around a PBH based on the distribution function $f(t,E,L^2)$. To do this, we first find the initial distribution function $f(t_i,E,L^2)$ in the regime when the average energy of thermal motion of DM particles is much less than the characteristic absolute value of their potential energy in the gravitational field of the PBH. In the work \cite{Adaetal19} it was found that for a standard neutralino such a situation is realized when the mass of the PBH is $M>10^{-6}M_\odot$. Note that in such a situation, the conserved energy $E$ can be assumed to be equal to the potential energy of the particle $-GMm/r_{\rm ta}$ at the turn-around point $r_{\rm ta}$, approximately coinciding (up to a factor of the order of one) with the radius of influence of the PBH at the time under consideration \cite{Ero16}
\begin{equation}
r_{\rm ta}(t)=(8GMt^2)^{1/3}.
\label{inflrad1}
\end{equation}
Thus, we neglect the radial component of the thermal velocity of the particles $v_r$. In contrast, the conserved angular momentum is determined by the orthogonal component of the thermal velocity $v_t$. In the spherically symmetric case, in the absence of other sources of angular momentum of particles, this orthogonal component is decisive for the distribution over $L^2$. Energy conservation follows from the fact that at $t\ll t_{eq}$ the total mass of DM in the density spike formed at the stage of radiation dominance is much less than the mass of the PBH. Therefore, the particles move in a constant gravitational potential of the PBH and their energy is conserved. The opposite situation is realized at the dust-like stage, when the total mass of the DM halo around a PBH exceeds the mass of the PBH itself, and in the growing halo the energy of the particles can change, however, in this paper we consider only the inner densest part of the halo, which is formed at radiation dominated stage. 

Let's consider the element of the DM mass in the density spike formed by the particles in the small intervals $dE$ and $dL$. Since the variables are the integrals of motion, then the redistribution of the particle by $dE$ and $dL$ does not occur, and when calculating the process of the density spike formation, we can limit ourselves to fixed intervals $dE$ and $dL$. The mass element $dM$ in these intervals in the already formed density spike is
\begin{equation}
\frac{dM}{dEdL}=4\pi m\int_{r_{\rm min}}^{r_{\rm max}}drr^2\frac{4\pi Lf(t,E,L^2)}{r^2v_r},
\label{dm1}
\end{equation}
where $r_{\rm min}$ and $r_{\rm max}$ are the roots of the equation $v_r=0$ for the given $E$ and $L$.

On the other hand, the same mass element in the same intervals $dE$ and $dL$ can be found at the moment of the expansion stop of the DM layer in the form
\begin{equation}
dM=\int d^3v m F(\vec v)4\pi r_{\rm ta}^2dr_{\rm ta}\rho(t_{\rm ta}),
\label{dm2}
\end{equation}
where the velocity distribution function is Maxwellian,
\begin{equation}
F(\vec v)=\frac{m^{3/2}}{(2\pi k_{\rm B}T)^{3/2}}e^{-mv^2/(2k_{\rm B}T)},
\label{maxwd}
\end{equation}
where $k_{\rm B}$ is the Boltzmann constant. The transition to the free-streaming regime occurs when particles are kinetically decoupled from radiation. Since the particles do not interact after the decoupling, their thermal distribution is preserved during the expansion of the universe even without thermalization. Free streaming does not affect the shape of the thermal distribution (\ref{maxwd}), but the temperature of the particle gas decreases as $T\propto1/a^2(t)$.

Decompose the velocity into radial and transverse parts $v^2=v_r^2+v_t^2$, then the angular momentum at the moment of the layer turnaround is expressed as $L=v_tr_{\rm ta}$. The temperature of the DM particle gas evolves as $T=T_dt_d/t$ \cite{Ero16}. Inside the radius $r_{\rm ta}(t_d)$ with the considered particle parameters, DM certainly annihilates, therefore we consider only the radii $r\gg r_{\rm ta}(t_d)$. Having integrated over all $v_r$, let's move in (\ref{dm2}) from the variables $dr_{\rm ta}$ and $dv_t$ to $dE$ and $dL$. After that, we equate the mass elements $dM$ from (\ref{dm1}) and (\ref{dm2}) and as a result we get
\begin{equation}
f(t,E,L^2)=\varkappa_1 \varepsilon^{1/4} e^{-l^2/l_m^2(\varepsilon)},
\label{fitog}
\end{equation}
where
\begin{equation}
\varkappa_1=\frac{\rho_d t_d^{1/2}}{2^{3/4}\pi^2 k_{\rm B}T_d(GMr^*)^{1/4}},
\end{equation}
and 
\begin{equation}
l_m^2(\varepsilon)=\varkappa_2\varepsilon^{-1/2}, \quad \varkappa_2=\frac{2^{5/2}k_{\rm B}T_d t_d}{m(GMr^*)^{1/2}}.
\label{lmeq}
\end{equation}
With the parameters under consideration, $\varkappa_1=4.6\times10^{-28}$~cm$^{-6}$~s$^3$, $\varkappa_2=4.5\times10^{-8}$.

For convenience, in (\ref{fitog}) we switched to the dimensionless units. We normalize the distances to the certain scale $r^*$, which is convenient to choose in the form $r^*=0.1r_{\rm eq}$, where
\begin{equation}
r_{\rm eq}=\left(\frac{3M}{4\pi\rho_{\rm eq}}\right)^{1/3}=0.15\left(\frac{M}{30M_\odot}\right)\mbox{~pc},
\end{equation}
where $\rho_{\rm eq}$ is the DM density at the moment of matter-radiation equality. The scale $r_{\rm eq}$ is the boundary of the density spike formed at the stage of radiation dominance, and the characteristic scale of the annihilation plateau is about 2-3 orders of magnitude smaller than $r_{\rm eq}$, therefore the specified choice of $r^*$ is convenient for numerical calculation. The energy and angular momentum are normalized as follows
\begin{equation}
\varepsilon=\frac{-Er^*}{GM}, \quad l=\frac{L}{(GMr^*)^{1/2}}.
\end{equation}

Using the distribution function (\ref{fitog}) one can find from (\ref{neq}) the DM density profile at the density peak. With good accuracy, it is approximated by a power-law dependence
\begin{equation}
\rho(r)=1.5\times10^{-19}\left(\frac{r}{r^*}\right)^\gamma\mbox{~g~cm$^{-3}$},
\label{rhouur}
\end{equation}
where $\gamma\approx -2.37$.
The density (\ref{rhouur}) at $r\simeq r^*$ is about 2.4 times less than follows from the expression
\begin{equation}
\rho(r)\simeq 1.5\rho_d\left(\frac{r}{r_d}\right)^{-9/4},
\end{equation}
obtained in \cite{Ire24} for the same radius, where $r_d$ is the turnaround radius at the moment $t_d$ of kinetic decoupling. On the contrary, our expression at $r\simeq r^*$ is about 3.7 times larger than the expression  
\begin{equation}
\rho(r)\simeq (\rho_{\rm eq}/2)t_{\rm eq}^{3/2}(2GM)^{3/4}r^{-9/4},
\end{equation}
given in \cite{Adaetal19}. Part of this difference can be explained by the choice of a coefficient of the order of one in the value of the turnaround radius (\ref{inflrad1}), numerically close to the radius of influence of  PBH. After the moment of equality, an extended DM halo is formed around a PBH by the mechanism of secondary accretion. However, this halo has virtually no effect on the structure of the central region itself, where the DM spike was previously formed, because the average density of the outer halo is several orders of magnitude lower than the DM density in the central density spike. Although the outer layers contain more mass, they have a much lower density and do not affect the central region.

As can be seen from (\ref{lmeq}), at masses $M\sim30M_\odot$, the DM particles have a very narrow distribution over $L$ near $L=0$, i.e. very narrow orbits with typical eccentricities $1-e\ll1$. This means that particles at a certain radius $r$ usually penetrate into the radii of $\ll r$ in their orbital motion, where they experience an increased rate of annihilation. As a result, even at a large distance from the annihilation plateau, the annihilation of particles on the elements of their orbits, which are close to the PBH, affects. On the other hand, particles can spend only a small part of the time inside the annihilation plateau compared to the orbital period. Therefore, their annihilation is weaker than assumed in the simplified approach, when the density of the annihilation plateau is calculated. In the next Section, we will calculate the effect of this annihilation near the PBH on the DM density change in the spike over long distances.


\section{Modification of the central density profile}
\label{itogsec}

Denoting $x=r/r^*$, $\nu=nr^{*3}$, $\phi=f(GMr^*)^{3/2}$, $\tau=t/t_0$, let's rewrite 
the equations (\ref{neq}) and (\ref{dfdtann2}) in the following dimensionless form, convenient for numerical calculation
\begin{equation}
\nu(\tau,x)=\frac{2\pi}{x}\int d\varepsilon'\int d{l'}^2\frac{\phi(\varepsilon',l')}{({\alpha'}^2-l^2)^{1/2}},
\label{neqbezr}
\end{equation}
\begin{equation}
\frac{\partial \ln\phi}{\partial \tau}=-\frac{1}{2\pi}k_0\varepsilon^{3/2}
\int\limits_{x_{\rm min}}^{x_{\rm max}}\frac{dx\, x\, \nu(\tau,x)}{(\alpha^2-l^2)^{1/2}}.
\label{dfdtann3}
\end{equation}
where 
\begin{equation}
\alpha=2x^2\left(-\varepsilon+\frac{1}{x}\right),
\end{equation}
$k_0=2^{5/2}\langle\sigma_{\rm ann} v\rangle t_0/r^{*3}\simeq7.2\times10^{-61}$ for the parameters used in this article. The integration limits are obtained by de-dimensioning the corresponding limits in the equations (\ref{neq}) and (\ref{dfdtann2}).

\begin{figure}[tbp]
\centering
\includegraphics[angle=0,width=0.49\textwidth]{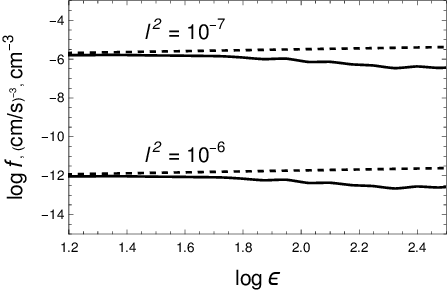}
\caption{Sections of the distribution function $f(t,E,L^2)$ depending on the dimensionless energy $\varepsilon$. The upper and lower dashed curves correspond to the initial distribution function for dimensionless angular momentum $l^2=10^{-7}$ and $l^2=10^{-6}$, correspondingly. The upper and lower solid curves show the modern $f(t_0,E,L^2)$ (after annihilation) for the same $l^2$. Small fluctuations on the lines are associated with computational errors.}
\label{gr2}
\end{figure}

In Fig.~\ref{gr2} sections of the initial distribution function $f(t_i,E,L^2)$ are shown. The distribution functions $f(t_0,E,L^2)$ at the present moment $t_0$, obtained from the numerical solution of the equations (\ref{neq}) and (\ref{dfdtann2}) with the initial distribution function (\ref{fitog}), are also presented.

\begin{figure}[tbp]
\centering
\includegraphics[angle=0,width=0.49\textwidth]{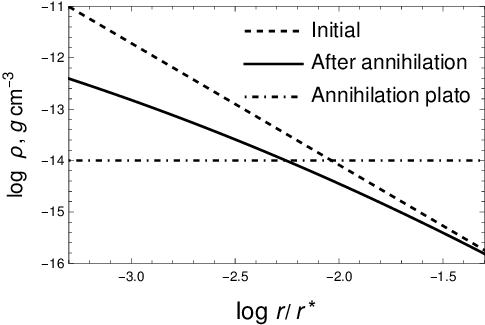}
\caption{DM density at the spike around a PBH, taking into account particle annihilation (solid curve) in comparison with the initial density profile (dashed curve) and the ``annihilation plateau'' $\rho\sim10^{-14}$~g~cm$^{-3}$ (dot-dashed curve).}
\label{gr3}
\end{figure}

Using the resulting modified distribution function, it is possible to calculate the final density in the spike, shown in Fig.~\ref{gr3} compared to the initial density. As can be seen from Fig.~\ref{gr3}, taking into account the orbital distribution, we no longer have a flat annihilation plateau. Instead, there is a smooth transition from the initial to the final profile. From the point of view of the annihilation signal, this means that the signal from the central region differs from the plateau case by a factor of the order of one, but to accurately calculate this difference, it will require calculating the structure of the central region of the density spike with greater computing power than was carried out in this paper. At small radii, instead of plateaus, an approximately power-law density distribution with an exponent of $\simeq-0.7$ occurs at the considered radii.

Taking into account the distribution of particle orbits weakens the limit associated with the annihilation plateau. But more fundamental restrictions for WIMPs follow from the effects of fermion degeneracy and from the conservation of phase volume according to Liouville's theorem (see, e.g., \cite{GorRub11}). But calculations show, that these restrictions for 70~GeV WIMPs give many orders of magnitude greater maximum density than the density of the annihilation plateau. Therefore, in the case under consideration, these restrictions are not important.

At a distance of more than a few gravitational radii, only the mass of the PBH is important, but not its angular momentum. Here, physics is practically Newtonian and the influence of angular momentum, described by the general theory of relativity, is already negligible. The angular momentum of a black hole is important only for the accretion of baryons. It determines the efficiency of energy release and, through this, the backreaction (the Eddington limit). For stellar-mass PBHs, the mass increase due to the accretion of baryon gas is insignificant, therefore, the mass increase will not greatly affect the central region of the DM density spike.


\section{Conclusion}

This paper presents the new method for calculating the DM density profile of the density spikes around PBHs and describes a method for investigating the modification of the density profile as a result of the DM annihilation. Of course, the latter effect only occurs in the case of annihilating particles such as WIMPs. In the calculations, it is important to take into account the distribution of DM particles in their orbits around a PBH due to the fact that a particle in its orbital motion passes through regions with different densities, as was indicated in \cite{Vas07,LacBea10}. Therefore, the decrease of particles at one distance from the PBH depends on their distribution in a wide range of distances along which the particle's orbit passes. This effect is especially important to take into account due to the fact that for sufficiently massive PBH, the orbits of particles are very narrow, almost radial, and they can approach close to the PBH, where the DM density is initially very high (before annihilation).

The numerical calculation performed gives index of the initial density profile $\gamma\approx-2.37$, slightly different from the value $-2.25$ obtained in the works \cite{Adaetal19} and \cite{Ire24}, where the angular momentum of particles associated with their thermal motion after kinetic decoupling was neglected. We assume that the slight difference is due to this residual thermal motion. After evolution during Hubble time, the slope of the profile in the center of the density spike changes and it acquires the index $\approx-0.7$ at the minimum radii achieved in our numerical calculation. Thus, over two orders of magnitude of radial distances, at which it was possible to trace the change in the density profile, a gentle slope appears instead of an annihilation plateau.

The method described in this paper for calculating the modification of the distribution function during annihilation can also be used for density peaks around astrophysical black holes, which was done in \cite{Vas07} for a power-law angular momentum distribution. For appropriate studies, one can apply the initial distribution functions found in \cite{GonSil99} for the case of an adiabatic increase in the DM density around a black hole. Our approach is limited to the case when the mass of the entire DM in the peak is much less than the mass of the black hole. Otherwise, with an increase in the mass of the DM halo, the gravitational potential will experience a change, which will lead to a redistribution of particles according to their energies $E$, although with spherical symmetry the angular momentum of the particles $L$ will not change. Due to the redistribution of particles by $E$, the equation (\ref{bolts}) can no longer be divided into two parts, and therefore the evolution of the distribution function will be under the joint influence of two processes, due to annihilation and changes in the gravitational potential.

If one a priori knows the density profile, then one can get a distribution function using the Eddington method. But the profile (\ref{rhouur}) was not known in advance. Suppose, however, that the density profile is known from somewhere. Then it should be noted that the original Eddington method, which allows solving the Abel integral equation, is applicable only to a distribution with isotropic velocity dispersion when the distribution function does not depend on angular momentum. In the case we are considering, this is not the case, our distribution function depends exponentially on angular momentum, and the orbits are almost radial. There is a generalization of the Eddington method with nonisotropic velocity dispersion \cite{Mer85}. However, in the process of annihilation, the distribution function evolves very inhomogeneously (most strongly at low energies and small angular momenta). For this reason, it is not possible to describe the distribution function  using the velocity dispersion anisotropy parameter introduced in \cite{Mer85}, and numerical calculation is required. 

This work is supported by the Russian Science Foundation grant 23-22-00013,

https://rscf.ru/en/project/23-22-00013/


\end{document}